\documentclass[showpacs,aps,twocolumn,nofootinbib,11]{revtex4}
\usepackage{amsmath}
\usepackage{epsfig}
\usepackage{subfigure}
\usepackage{float}
\usepackage{longtable}
\usepackage{times}
\usepackage{txfonts}
\usepackage{color}
\usepackage{enumitem}

 \usepackage{ifpdf}
 \ifpdf
 \usepackage[pdftex]{hyperref}
 \else
 \usepackage[hypertex]{hyperref}
 \fi

 \hypersetup{
   pdftitle={},%
   pdfauthor={},%
   pdfsubject={},%
   pdfkeywords={},%
   pdfstartview={},%
   bookmarksopen=true, breaklinks=true, debug=true, %
   colorlinks=true, linkcolor=red, citecolor=blue, urlcolor=blue
 }

\newcommand{\beq}{\begin{eqnarray}}
\newcommand{\eeq}{\end{eqnarray}}
\newcommand{\be}{\begin{eqnarray*}}
\newcommand{\ee}{\end{eqnarray*}}

\newcommand{\ie}{{\it i.e.}}

\newcommand{\ce}[1]{Eq.~\eqref{#1}}

\newcommand{\cf}[1]{{Fig.~\ref{#1}}}

\def\lsim{\raise0.3ex\hbox{$<$\kern-0.75em\raise-1.1ex\hbox{$\sim$}}}
\def\gsim{\raise0.3ex\hbox{$>$\kern-0.75em\raise-1.1ex\hbox{$\sim$}}}

\def\d+Au  {$d$Au}
\def\d+Aum  {d\mathrm{Au}}
\def\pPb {$p$Pb}

\def\pp   {$pp$}
\def\pA   {$pA$}
\def\AA   {$AA$}

\def\PbPb {PbPb}

\def\sqrtsNN {\mbox{$\sqrt{s_{NN}}$}}
\def\Npart   {\mbox{$N_{\rm part}$}}

\def\RpPboneS    {\mbox{$R_{p\rm Pb}^{\Upsilon(1S)}$}}
\def\RPbPbnS    {\mbox{$R_{PbPb}^{\Upsilon(nS)}$}}

\def\beq     {\begin{equation}}
\def\eeq     {\end{equation}}

\long\def\symbolfootnote[#1]#2{\begingroup%
\def\thefootnote{\fnsymbol{footnote}}\footnote[#1]{#2}\endgroup}

\begin{document}

\title{Is bottomonium suppression in proton-nucleus and nucleus-nucleus collisions at LHC energies due to the same effects?}

\author{E. G.~Ferreiro}
\affiliation{Laboratoire Leprince-Ringuet, Ecole polytechnique, CNRS/IN2P3, Universit\'e Paris-Saclay, Palaiseau, France}
\affiliation{Departamento de F{\'\i}sica de Part{\'\i}culas and IGFAE, Universidade de Santiago de Compostela, 15782 Santiago de Compostela, Spain}

\author{J.P. Lansberg}
\affiliation{IPNO, Universit\'e Paris-Saclay, Univ. Paris-Sud,  CNRS/IN2P3, F-91406, Orsay, France }

\begin{abstract}
We show that we can reproduce all the features of bottomonium suppression in both proton-nucleus and nucleus-nucleus collisions at LHC energies in a comover-interaction picture. 
For each collisions system, we use the measured {\it relative} suppression of the excited $\Upsilon$(2S) and $\Upsilon$(3S) states to $\Upsilon$(1S) by ATLAS and CMS to parametrise the scattering cross sections of all $S$- and $P$-wave bottomonia with the comoving particles 
created during the collisions. In addition to a single nonpertubative parameter, these cross sections depend on the momentum distribution of these comovers which we found to be the same for proton-nucleus and nucleus-nucleus collisions as well as for partonic and hadronic comovers. 
Moreover, we can also reproduce the {\it absolute} suppresion rates measured by ALICE, ATLAS, CMS and LHCb when the nuclear modifications of the parton densities are taken into account. 
\end{abstract}

\pacs{13.85.Ni,14.40.Pq,21.65.Jk,25.75.Dw}
\maketitle

{\it Introduction. ---}
By the end of the year 2013, following the LHC \pPb\ run at $\sqrtsNN=5$ TeV, the CMS Collaboration probably reported  the most 
unexpected observation of the LHC heavy-ion programme: the excited $\Upsilon$(2S) and $\Upsilon$(3S) states were experiencing 
more suppression than the lower $\Upsilon$(1S) state~\cite{Chatrchyan:2013nza}. At low collision energies, 
such a relative suppression would naturally follow from {\it final}-state interactions with the remnants of the colliding lead nucleus. Yet, at LHC energies,
the produced $b\bar b$ pair does not have the time to evolve into any physical state when it escapes the nuclear matter contained
in the colliding nucleus. As such, 
one cannot invoke this mechanism to explain the relative suppression observed by CMS, neither can one invoke {\it initial}-state effects
such as the modification of parton flux~\cite{Kovarik:2015cma,Owens:2012bv, deFlorian:2011fp,Eskola:2009uj,Hirai:2007sx}
or coherent energy loss~\cite{Arleo:2014oha}, which are known to have a similar impact on the different bottomonia~\cite{Ferreiro:2011xy}. Very recently, the ATLAS collaboration confirmed~\cite{Aaboud:2017cif} the observation of these relative suppressions with very similar magnitudes.

Not only was this result totally unforeseen, but it casted serious doubts on the conventional interpretation of the
relative suppression of bottomonium earlier observed by CMS in lead-lead collisions~\cite{Chatrchyan:2011pe,Chatrchyan:2012lxa}: 
the excited $\Upsilon$(2S) and $\Upsilon$(3S) states are larger, less tightly bound and thus suffer more from the colour screening
and related effects in the quark-gluon plasma (QGP). Quantitatively, if the nuclear effects responsible for the unexpected relative suppression in \pPb\
collisions add up linearly in \PbPb\ collisions in each nucleus --or equally speaking if they factorise-- 
they are expected to be responsible for half of the \PbPb\ relative suppression~\cite{Lansberg:2015uxa}.
If they were subtracted, the QGP effects would thus be half of what they have been so far believed to be.

In this work, we attempt to explain these relative suppression in \pPb\ and \PbPb\ collisions 
altogether by assuming that the bottomonia are broken by collisions with comoving particles
--\ie\ particles with similar rapidities-- and whose density is directly connected
to the particle-multiplicity measured at that rapidity for the corresponding colliding system.
In such a scenario, an increase in the colliding energy has the opposite effect than 
for the suppression by the nucleus remnants. Instead of decreasing because of color transparency, or
because the propagating pair can only interact in smaller amounts of time, it increases because the 
number of produced particles from a given \pPb\ collision increases with energy. So does the number of comoving
particles along with the heavy-quark pair. Another important feature of this assumption is that
the heavy-quark pairs have reached --in their rest frame-- a physical state after a
fraction of a femtometer. The $\Upsilon$(1S),  $\Upsilon$(2S) and $\Upsilon$(3S) states
then interact with very different probabilities with these comoving particles, which provides a very natural explanation
for the observed relative suppression. 

As aforementioned, nobody expected this observation and, as for now, no other effect 
have been proposed to explain it apart from suggesting the creation of a ``hot'' medium in these high-energy
proton-nucleus ($pA$) collisions. None of the known ``cold" nuclear-matter effects generate a relative
suppresion. It is thus in fact the ideal observable to fix the comover-bottomonium cross sections, which 
are the only new phenomenological quantities entering our study. Surprisingly, the CIM has never been applied
to the bottomonia.

We in fact go one step further by proposing an improved version of the well-established comover interaction model (CIM) \cite{Capella:1996va,Armesto:1997sa,Armesto:1998rc,Capella:2000zp,Capella:2005cn,Capella:2007zz,Capella:2007jv,Ferreiro:2012rq}, 
already successfully applied to explain a similar unexpected suppression of excited
{\it charmonia}~\cite{Ferreiro:2014bia}. Indeed, instead of independently fixing the cross sections
state by state, we propose a generic formula for all the quarkonia states and suggest a connection
with the momentum distribution of the comovers in the transverse plane, thus with an effective
temperature ($T_{\rm eff}$) of the comovers. With such an approach, we are able to propose a clear
benchmark between $pA$ and $AA$ collisions under the CIM paradigm.

As we shall see, the approach is particularly successful: 
\begin{enumerate}[label=(\roman*),topsep=0pt,
    itemsep=-1ex,
    partopsep=1ex,
    parsep=1ex,
    leftmargin=0ex,
    itemindent=4ex]
\item the interaction strengths between the 
bottomonia and the comovers needed to reproduce the 
\pPb\ data 
follow a simple pattern in terms of 
the size and the binding
energy -- both calculable with a simple Schr\"odinger equation-- of all the bottomonium states, which renders our set-up {\it predictive}; 
\item even more striking,
the entire {\it relative} suppression observed in \PbPb\ collisions is accounted
by scatterings with comovers with remarkably similar interaction strength as for the
\pPb\ data; 
\item the {\it absolute} magnitude of the $\Upsilon$ suppression
in \pPb\ and \PbPb\ collisions 
is also very well reproduced
up to the uncertainties in the nuclear modification of the gluon densities.
\end{enumerate}
 
Overall, as we will show, all the LHC \pPb\ and \PbPb\ data can be reproduced with merely two parameters.

{\it The Comover Interaction Model. ---} 
Within this framework, the quarkonia are suppressed by the interaction 
with the comoving medium, constituted by particles with similar rapidities. 
At a time $\tau$,
the rate equation that governs the density of quarkonium at a given transverse coordinate $s$ and rapidity~$y$ for a collision of impact parameter $b$, $\rho^{\Upsilon}(b,s,y)$, obeys the expression
\beq
\label{eq:comovrateeq}
\tau \frac{\mbox{d} \rho^{\Upsilon}}{\mbox{d} \tau} \, \left( b,s,y \right)
\;=\; -\sigma^{\rm co -\Upsilon}\; \rho^{\rm co }(b,s,y)\; \rho^{\Upsilon}(b,s,y) \;,
\eeq
where $\sigma^{\rm co -\Upsilon}$ is the cross section of bottomonium dissociation
due to interactions with the comoving medium characterized by transverse density~$\rho^{\rm co }(b,s,y)$
at $\tau_i$.
By integrating this equation from $\tau_i$ to $\tau_f$, 
one obtains the survival probability $S^{\rm co }_{\Upsilon}(b,s,y)$  of a $\Upsilon$ 
interacting with comovers:
$S^{\rm co }_{\Upsilon}(b,s,y)  \;=\; \exp \Big\{-\sigma^{\rm co -\Upsilon}
  \, \rho^{\rm co }(b,s,y)\, \ln
\Big({\rho^{\rm co }(b,s,y)}/{\rho_{pp} (y)}\Big) \Big\}
$, 
where the argument of the logarithm comes from $\tau_f/\tau_i$ converted in ratios of densities where
we assumed that the interaction stops at $\tau_f$ when the densities
have diluted, reaching the value of the \pp\ density at the same energy and rapidity, $\rho_{pp}$.

In order to compute $S^{\rm co }_{\Upsilon}$, 
the comover density, $\rho^{\rm co }$,  is mandatory. It is directly obtained from
the particle multiplicity measured at the rapidity $y$	 for the corresponding 
colliding system.

For \pA\ collisions, it is most natural to take the medium as made of pions. 
Nevertheless, we will show later that the nature of this medium --partonic or hadronic-- does not change our results which is one
of the important findings of our study.

As can be seen from \ce{eq:comovrateeq}, the main ingredient driving the abundance of a given bottomonium is its interaction cross section with the comovers, $\sigma^{\rm co -\Upsilon}$.  In our previous works on charmonia, these were obtained from fits
to low-energy $AA$ data~\cite{Armesto:1997sa}, $\sigma^{\rm co -J/\psi}=0.65$ mb and $\sigma^{\rm co -\psi(2S)}=6$ mb. Such 
--purely phenomenological-- cross sections in fact would result  from the convolution of the comover momentum distribution in the transverse plane and 
the momentum-dependent comover-quarkonium cross section. As such they may slightly depend on the collision energy via 
a change of the comover momentum distribution.
Yet, these values were successfully applied at higher energies to reproduce~\cite{Ferreiro:2014bia} $J/\psi$ and $\psi$(2S) $pA$ data at RHIC and the LHC as well as $AA$ data accounting for the recombination of charm quarks~\cite{Capella:2007jv,Ferreiro:2012rq}. 

One can not follow the same approach for $\Upsilon(nS)$ since no $AA$ relative-suppression bottomonium data exist at low energies 
and, in fact, the CIM was never applied to bottomonia before. In addition, the bottomonium family is richer
with at least 6 phenomelogical cross sections to be considered in a full computation. 
We have adopted another strategy by going to a slightly more microscopic level 
accounting for the momentum distribution of the comover-quarkonium cross section and 
that of the comovers in the transverse plane. This in fact allowed
us to reduce the degrees of freedom  of our modeling to the introduction of essentially 2 parameters, yet applicable to the entire bottomonium family 
and allowing us to investigate the nature the comovers (gluons or pions).  
To do so we assumed that:
\begin{enumerate}[label=(\roman*),topsep=0pt,
    itemsep=-1ex,
    partopsep=1ex,
    parsep=1ex,
    leftmargin=0ex,
    itemindent=4ex]
\item
the thresholds approximately follow from the mass differences 
between the quarkonium and the lightest open beauty hadron pair, taking into account the comover mass;
\item 
 away from the threshold, the cross section should scale like the geometrical cross section $\sigma^{\cal Q}_{\rm geo} \simeq \pi r_{\cal Q}^2$,
where $r_{\cal Q}$ the quarkonium Bohr radius. It can be evaluated by solving the Schr\"{o}dinger equation with a well-choosen potential reproducing the quarkonium spectroscopy~\cite{Satz:2005hx}.

\end{enumerate}

Our parametrization of the energy dependence thus simply amounts to interpolating from $\sigma^{\rm co-{\cal Q}}(E^{\rm co})=0$ at threshold up to $\sigma^{\rm co-{\cal Q}}(E^{\rm co})=\sigma^Q_{\rm geo}$ 
away from threshold but with the {\it same} dependence for {\it all} the states. It reads 
$\sigma^{\rm co-{\cal Q}}(E^{\rm co}) =\sigma^Q_{\rm geo}  (1-{E^Q_{\rm thr}}/{E^{\rm co }})^n$
where $E^Q_{\rm thr}$ corresponds to the threshold energy to break the quarkonium bound state and
$E^{\rm co}=\sqrt{p^2+m_{\rm co}^2}$ is the energy of the comover in the quarkonium rest frame.
The first parameter of our modeling, $n$, characterizes how quick the cross section approaches
the geometrical cross section. 
Attempts to compute this energy dependence, using the multipole expansion in perturbative QCD at LO~\cite{Bhanot:1979vb,Kharzeev:1994pz,Satz:2005hx}, would suggest that $n$ is close to 4 for pion comovers by making the strong assumption that the scattering is initiated by gluons inside these pions.  
Hadronic models which take into account non-perturbative effects and thus most likely provide a better description of the physics at work \cite{Rapp:2008tf} 
show a different energy dependence.
 It effectively corresponds to smaller $n$ \cite{Martins:1994hd}.
As such, we will consider $n$ varying from 0.5 to 2.
In fact, the discrepancies existing
between the aforementioned LO QCD results and these hadronic calculations are partly due to 
large higher order correction near the threshold \cite{Song:2005yd}. 

As for the momentum distribution of the comovers in the transverse plane, we simply take a 
Bose-Einstein distribution
${\cal P}(E^{\rm co };T_{\rm eff}) \propto 1/(e^{E^{\rm co }/T_{\rm eff}}-1)$
which introduces our second parameters, namely an effective temperature of these comovers.
Our fits will thus simply amount to determine the best value $T_{\rm eff}$ for fixed values of $n$ in the aforementioned ranges.

{\it Fitting the data.---} 
In order to proceed with the fit, it is mandatory to take into account the feed-down contributions.
In fact, the observed $\Upsilon$(nS) yields contain contributions from decays of heavier bottomonium
states and, thus, the measured suppression can be affected by the dissociation of these states.
This feed-down contribution to the $\Upsilon$(1S) state is usually asumed to be on the order of 50\%, according to the CDF measurements at $p_T > 8$ GeV \cite{Affolder:1999wm}.
However, following the LHCb data extending to lower $p_T$~\cite{Aaij:2014caa}, this assumption needs to be revisited, in particular for $p_T$-integrated results. In such a case, the feed-down fractions for the $\Upsilon$(1S) are rather: 70\% of direct $\Upsilon$(1S), 8\% from $\Upsilon$(2S) decay, 1\% from $\Upsilon$(3S),
15\% from $\chi_{b}$(1P), 5\% from $\chi_{b}$(2P) and 1\% from $\chi_{b}$(3P), 
while for the $\Upsilon$(2S) the different contributions would be: 63\% direct $\Upsilon$(2S), 4\% of $\Upsilon$(3S), 30\% of $\chi_{b}$(2P)  and 3\% of  $\chi_{b}$(3P)~\cite{Andronic:2015wma}.
Note also that for the  $\Upsilon$(3S), 40\% of the contribution is expected from decays of  $\chi_{b}$(3P). However, in this case, the measurement was done for $p_T > 20$ GeV.
These fractions remain partly extrapolated resulting in possible additional uncertainties to be taken into account. We have thus varied
the feed-down fractions for two limiting cases : 80\% of direct $\Upsilon$(1S) and 50\% of direct $\Upsilon$(3S), 60\% of direct $\Upsilon$(1S) and 70\% of direct $\Upsilon$(3S), leaving the other ones unchanged. This however does induce changes which are not significant in view of the current experimental uncertainties.

As announced, we performed our fit on {\it relative} --minimum bias-- nuclear suppression factors. For \pPb\ collisions, we have used the CMS \cite{Chatrchyan:2013nza} and ATLAS \cite{Aaboud:2017cif} data. For \PbPb\ collisions, we have used the CMS data at 2.76 TeV~\cite{Chatrchyan:2012lxa} and at 5.02 TeV~\cite{Sirunyan:2017lzi}. For both these \pPb\ and \PbPb\ cases, we performed the fit of $T_{\rm eff}$ for different values of $n$ with both gluon or pion comovers. Our results are depicted on \cf{fig:figTvsn}. The resulting uncertainty on $T_{\rm eff}$ is from the experimental uncertainty. Up to this uncertainty, all  the combinations yield to the same couple $(n,T_{\rm eff})$ with $T_{\rm eff}$ in the range 200 to 300 MeV for our assumed range for $n$. Our fits are equally good with $\chi^2_{\rm d.o.f.}$ ranging, for \pPb\ data, from 1.0 to 1.4 and, for \PbPb\ data, from 1.4 to 2.0. 

\begin{figure}[thb]
\hspace{-0.5cm}
\includegraphics[width=0.9\linewidth]{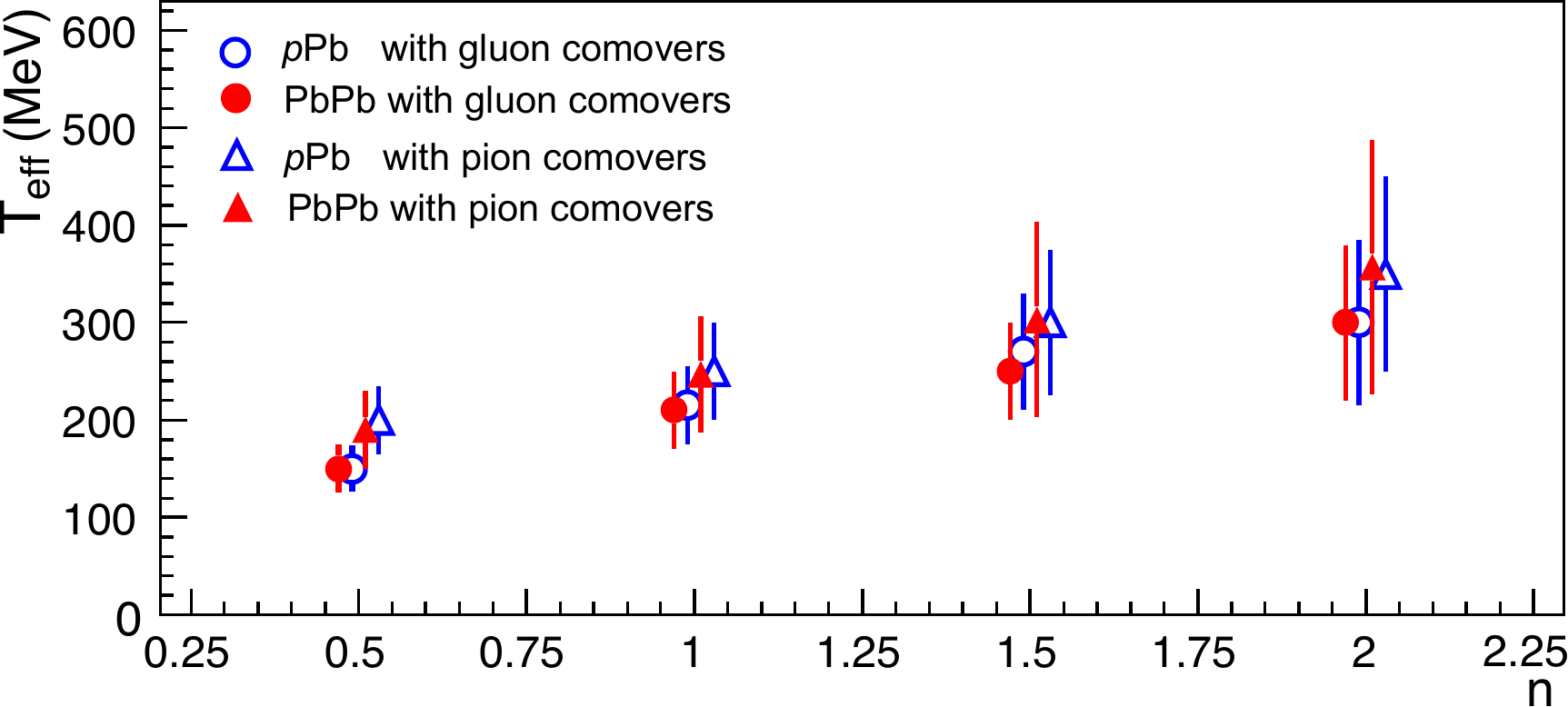}
\vskip -0.2cm
\caption{\label{fig:figTvsn}
Fitted $T_{\rm eff}$ considering  pion (triangles) or gluon (circles) comovers from our fits to  \pPb\ (empty blue) and \PbPb\ (filled red) data
for different $n$ from 0.5 to 2. [The points have been horizontally shifted for readibility.]
}
\end{figure}

Thus, we are confronted to the following quasi equiprobable possibilities:\\
Case I: The medium is of hadronic nature in  \pPb\ collisions, while it is gluonic in \PbPb\ collisions.\\
Case II: Both in \pPb\ and \PbPb\ collisions, the medium is made of hadrons, \ie\ the comovers can be identified with  pions.\\
Case III: Both in \pPb\ and \PbPb\ collisions, the medium is made of partons, \ie\ the comovers can be identified with gluons.\\
Case IV: The medium is of gluonic nature in  \pPb\ collisions, while it is hadronic in \PbPb\ collisions.\\
Case I is the most common expectation. The relevant d.o.f. are hadrons in \pPb\ collisions where the QGP is not produced whereas the
gluons become  relevant in the hotter \PbPb\ environment with the presence of QGP. Case II is the usual interpretation of historical CIM studies for which the gluon d.o.f. do not appear to be relevant. At SPS energies, it is a reasonable assumption. At the LHC, it is more thought-provoking, yet compatible with the observed bottomonium suppression at the LHC. It can also be understood in the sense that the melting temperature of the $\Upsilon$(1S) and $\Upsilon$(2S) is too high to be observed and the $\Upsilon(3S)$ is fragile enough to be entirely broken by hadrons. Case III amounts to say that gluons are the relevant d.o.f. to account for bottomonium suppression in both \pPb\ and in \PbPb\ collisions. One could thus say that a QGP-like medium is formed following \pPb\ collisions at LHC energies. Case IV is admittedly an unexpected situation.

In what follows, our results will be shown for $n=1$ and $T_{\rm eff}=250 \pm 50$ MeV. 
As what regards the $\Upsilon$-comover cross sections, for an exponent $n=1$ and taking into account the uncertainty in the temperature, 
$T_{\rm eff}=250 \pm 50$~MeV, we have from $\sigma^{\rm co -\Upsilon{\rm (1S)}}=0.02^{+0.020}_{-0.010}$ mb for the most tightly bound state $\Upsilon{\rm (1S)}$, compatible with no suppression of direct  $\Upsilon{\rm (1S)}$, to $\sigma^{\rm co -\chi_b{\rm (3P)}}=12.55^{+1.53}_{-1.88}$ mb for the loosely bound $\chi_{b}$(3P) states in the hadronic case.
Looking at these cross sections allows us to better understand the small impact of considering gluon or pion comovers. In fact, the mass effects only matter for $\chi_{b}$(3P) states altering their interaction cross section by 25\%
which however does not induce visibly different suppression effects. Indeed, for such large cross sections, the obtained suppression is already maximal for minimum bias collisions.

{\it Relative nuclear modification factors.---}
The resulting relative nuclear modification factors of the excited bottomonium states to 
their ground state in \pPb\ collisions at 5.02 TeV are presented in Table~\ref{tab:pPb} 
and compared to the CMS \cite{Chatrchyan:2013nza} and ATLAS \cite{Aaboud:2017cif} experimental data. 
We note that the central values of the data tend to indicate a slightly stronger suppression that
our results. We however recall that so far no other model could explain this relative suppression in 
\pPb\ collisions.

\begin{table}[hbt!]
\begin{center}\setlength{\arrayrulewidth}{1pt}
\caption{$\Upsilon$ \pPb\ at 5.02 TeV}\label{tab:pPb}
\begin{tabular}{cccc}
\hline\hline
& CIM & Exp\\
\hline
 & $-1.93 < y < 1.93$ & CMS data\\
 $\Upsilon{\rm (2S)}/ \Upsilon{\rm (1S)}$ &0.91 $\pm$ 0.03 & 0.83 $\pm$ 0.05 (stat.) $\pm$ 0.05 (syst.)\\
  $\Upsilon{\rm (3S)}/ \Upsilon{\rm (1S)}$  & 0.72 $\pm$ 0.02  & 0.71 $\pm$ 0.08 (stat.) $\pm$ 0.09 (syst.)\\
 & $-2.0 < y < 1.5$ & ATLAS data\\
  $\Upsilon{\rm (2S)}/ \Upsilon{\rm (1S)}$ &0.90 $\pm$ 0.03 & 0.76 $\pm$ 0.07 (stat.) $\pm$ 0.05 (syst.)\\
  $\Upsilon{\rm (3S)}/\Upsilon{\rm (1S)}$  & 0.71 $\pm$ 0.02  & 0.64 $\pm$ 0.14 (stat.) $\pm$ 0.06 (syst.)\\
 \hline\hline
\end{tabular}
\end{center}\vspace*{-0.5cm}
\end{table}

In \PbPb\ collisions, besides the minimum-bias values which we used in our fits, CMS reported on 
the centrality dependence of the relative suppression of $\Upsilon$(nS) at 2.76 and 5.02~TeV~\cite{Chatrchyan:2012lxa,Sirunyan:2017lzi}.
\cf{fig:figdoubleratioPbPb} shows our results along with the CMS points. The agreement is very good at 2.76 TeV, a bit less for $\Upsilon{\rm (2S)}/\Upsilon{\rm (1S)}$ at 5.02 TeV.

\begin{figure*}[htb!]
\hspace{-0.2cm}
\includegraphics[width=0.95\linewidth]{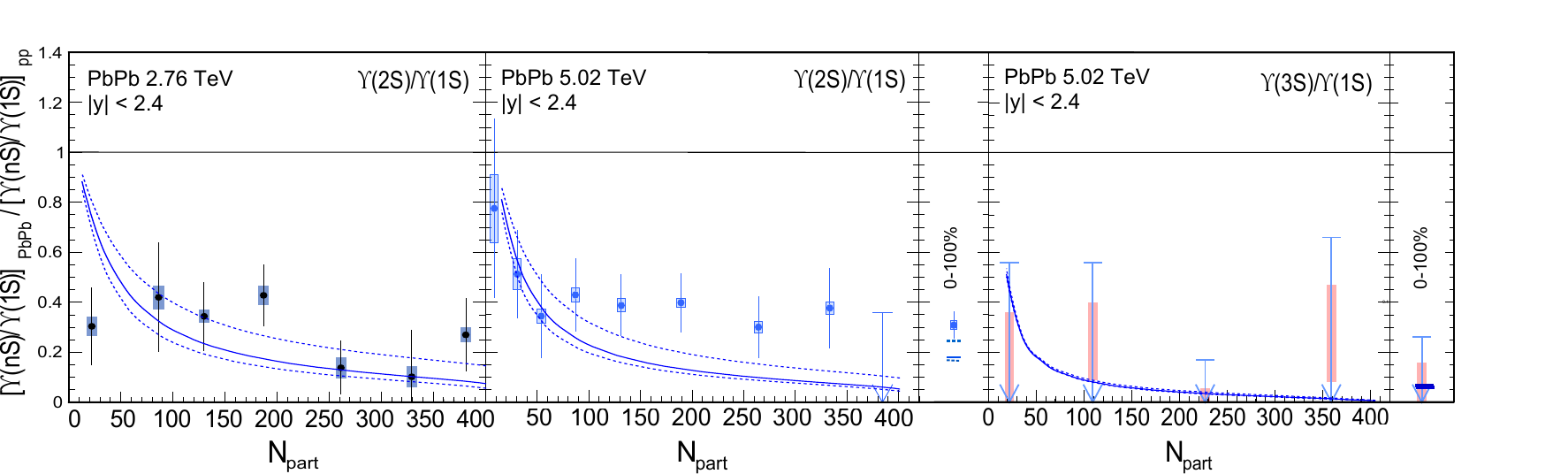}
\vskip -0.2cm
\caption{\label{fig:figdoubleratioPbPb}
The double ratio for $\Upsilon$(2S) over $\Upsilon$(1S) at 2.76 and 5.02 TeV and $\Upsilon$(3S) over $\Upsilon$(1S) at 5.02 TeV
as a function of \Npart\ obtained from CIM
compared to the CMS data at 2.76 TeV \cite{Chatrchyan:2012lxa} and 5.02 TeV~\cite{Sirunyan:2017lzi}.
The dashed line depicts the uncertainty from the fit of $\sigma^{co-\Upsilon}$.
}
\end{figure*}
{\it Absolute nuclear modification factors.---}
Having fixed the parameters of our approach with the relative suppression measurements, we can now address
the absolute suppression of each measured states. 
However, when addressing the absolute $\Upsilon$ suppression,
other nuclear effects, which cancel in the double ratio of the excited-to-ground state suppression,
do not cancel anymore. At LHC energies, the main one seems to be~\cite{Kusina:2017gkz} the nuclear modification 
of the Parton Distribution Function (PDFs). It is easily accounted for by using available global nPDF fits with uncertainties~\cite{Eskola:2009uj,Kovarik:2015cma,Eskola:2016oht,deFlorian:2011fp}.
In particular, we used nCTEQ15 which describes very well the suppression of open charm in \pPb\ collisions at the LHC~\cite{Kusina:2017gkz}. We also note that the central value of the nCTEQ15 fit is compatible with the one of EPS09LO  \cite{Eskola:2009uj} previously used in \cite{Ferreiro:2014bia}.

\begin{figure}[hbt!]
\hspace{-0.3cm}
\includegraphics[width=0.8\linewidth]{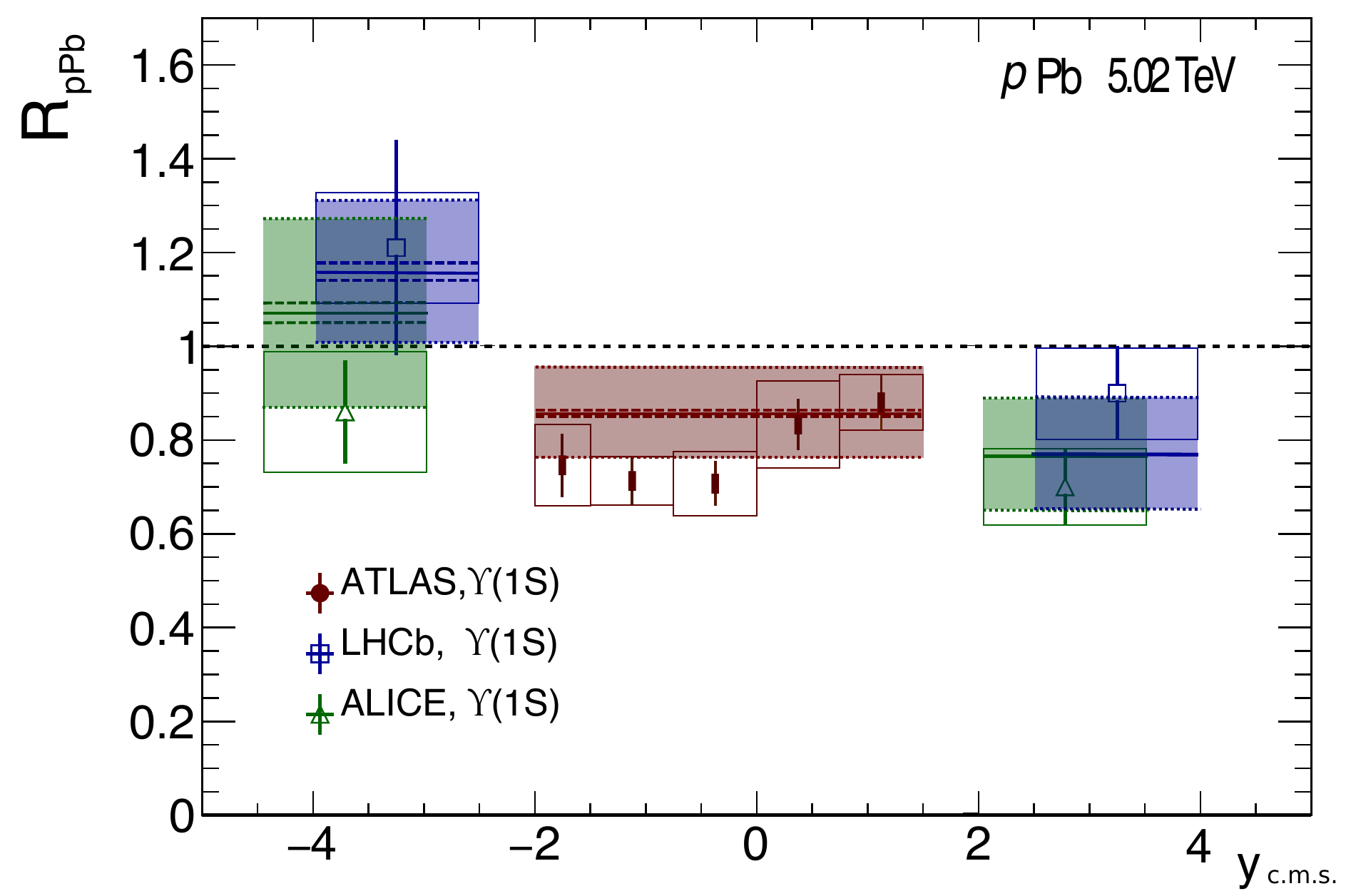}
\vskip -0.2cm
\caption{\label{fig:fig1SpPb}
The 
\RpPboneS\ vs rapidity compared to LHC data \cite{Aaij:2014mza,Abelev:2014oea,Aaboud:2017cif}. 
The uncertainty from the $\sigma^{co-\Upsilon}$ fit (dashed line) and from nCTEQ15 shadowing (dotted line along with the colored band) are shown separately.
}
\end{figure}
Let us first start with the $\Upsilon$(1S) case in \pPb\ collisions.
\cf{fig:fig1SpPb} shows the nuclear modification factor 
\RpPboneS\ vs rapidity at $\sqrt{s}=5.02$~TeV compared to available experimental 
data~\cite{Aaij:2014mza,Abelev:2014oea,Aaboud:2017cif} from ALICE, ATLAS and LHCb. The agreement is overall very good 
and the additional effect of the CIM is to damp down the antishadowing peak in the backward rapidity region which brings the
theory closer to the central value of ALICE.

We then compare our results with \PbPb\ data whose centrality dependence has also been measured. To address this dependence,
we parameterized the impact-parameter dependence of the nPDF as in~\cite{Ferreiro:2014bia}.
Our results for the 3 $\Upsilon$ states at 2.76 (5.02) TeV are shown in \cf{fig:figRPbPb} up (down) and compared to the CMS data \cite{Khachatryan:2016xxp,CMS:2017ucd}.
A good agreement is obtained in the 3 cases with the same parameters used to reproduce the relative suppression. 
\begin{figure}[hbt!]
\hspace{-0.3cm}
\includegraphics[width=0.7\linewidth]{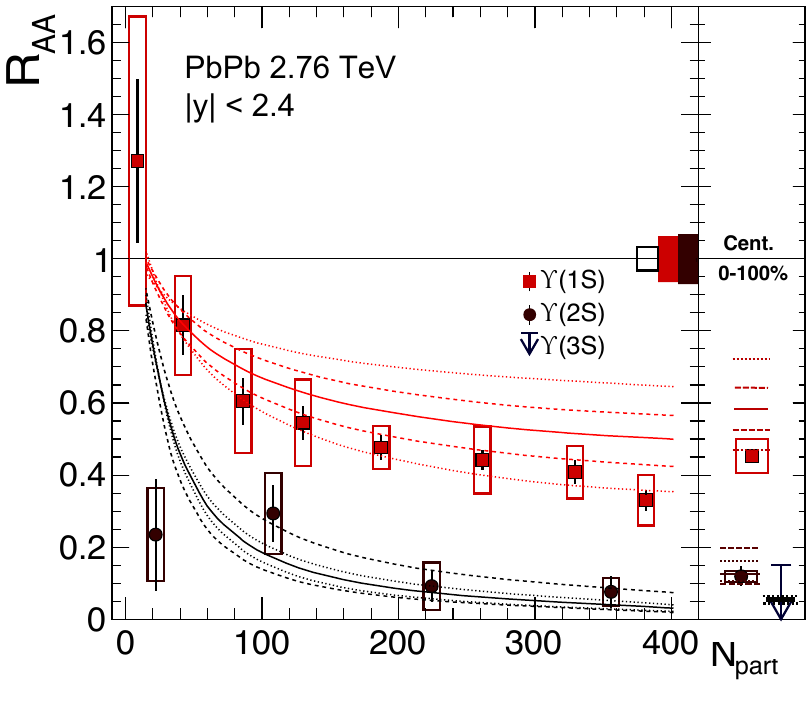}
\includegraphics[width=0.7\linewidth]{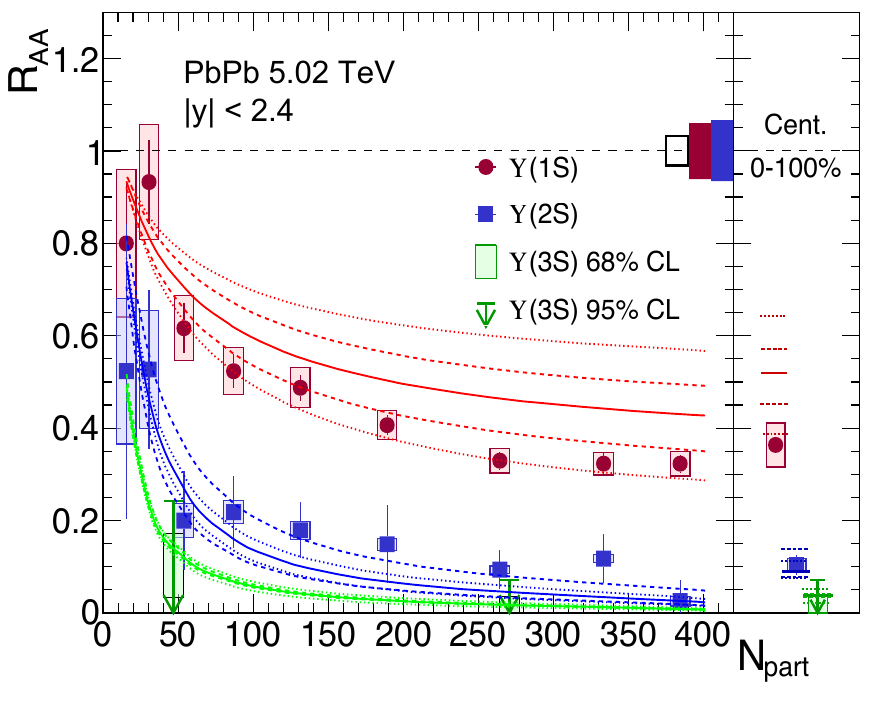}
\vskip -0.2cm
\caption{\label{fig:figRPbPb} 
\RPbPbnS vs \Npart\
compared to the CMS data at 2.76 TeV \cite{Khachatryan:2016xxp} and 5.02 TeV~\cite{CMS:2017ucd}.
The uncertainty from the fit of $\sigma^{co-\Upsilon}$ (dashed line) and from the nCTEQ15 shadowing (dotted line) are shown separately.
}
\vskip -0.5cm
\end{figure}

{\it Conclusions.---} 
In this work, we have addressed the puzzle of the relative suppression of the excited 
bottomonium states as compared to their ground state in \pPb\ collisions.
In the absence of any other explanation, we have assumed that
the reinteraction with comovers explains it all. This motivated us to revisit the CIM and to propose 
a generic formula for all the bottomonium states interpolating from the absence of interaction at threshold
up to the geometrical one for increasing comover-quarkonium relative momenta in the transverse plane. Taking into account the 
momentum distribution of the comovers with a Bose-Einstein distribution, we could further
study the impact of considering either massive pion or massless gluon comovers and 
investigate an effective temperature of the comovers as probed by the quarkonia.
This allowed us to fit the CMS and ATLAS \pPb\ double ratios with only 2 parameters 
accounting for all the comover-bottomonium interaction cross sections.
With the same setup, an independent fit of the corresponding \PbPb\ CMS data yielded
similar fit parameter values, thus hinting at a similar momentum
distribution of these comovers in the environment created by \pA\ and \AA\ collisions.
This is admittedly an unexpected and very interesting observations.

We further backed up our investigations by noting that our approach correctly predicted the absolute 
$\Upsilon$ suppression in both \pPb\ and  \PbPb\ collisions when combined with
nCTEQ15 shadowing without the need to invoke any other phenomena.

{\it Acknowledgements.---} We would like to thank Francois Arleo, Alfons Capella, Olivier Drapier and Frederic Fleuret for stimulating and useful discussions. E.G.F. thanks Laboratoire LePrince-Ringuet de l'Ecole Polytechnique and l'Universit\'e Paris-Saclay for their hospitality and financial support in the framework of Jean D'Alembert fellowship during the completion of this work, and to the Ministerio de Ciencia e Innovacion of Spain under project
FPA2014-58293-C2-1-P for financial support. The work of J.P.L. is supported in part by the French CNRS via the grant
PICS-07920 "Excitonium".

\bibliographystyle{utphys}

\bibliography{Upsi-Comover-120418}

\end{document}